\documentclass[preprint2]{aastex}

\begin{document}

\title{{\it Chandra} and {\it ASCA} X-ray Observations of the Radio
Supernova SN1979C IN NGC 4321}

\author{A. Ray \altaffilmark{1}}

\affil{Tata Institute of Fundamental Research, Bombay 400 005, India}

\author{R. Petre}

\affil{Lab. of High Energy Astrophysics, NASA/Goddard Space Flight Center,
Greenbelt, MD 20771, USA}

\author{E. M. Schlegel}

\affil{Harvard-Smithsonian Center for Astrophysics, 60 Garden Street, Cambridge, MA 02138, USA}

\altaffiltext{1}{formerly NAS-NRC Senior Research Associate at
NASA/GSFC, Greenbelt, MD 20771; E-mail: akr@tifr.res.in}


\begin{abstract}

We report on the X-ray observation of the radio selected supernova
SN1979C carried out with {\it ASCA} in 1997 December and
serendipitously available from a {\it Chandra} Guaranteed Time
Observation in 1999 November. The supernova, of type SN II-Linear (SN
II$_{\rm L}$), was first observed in the optical and occurred in the
weakly barred, almost face on spiral galaxy NGC 4321 (M100). The
galaxy, a member of the Virgo S cluster, is at a distance of 17.1 Mpc,
and contains at least three other supernovae discovered in this
century. The useful exposure time was $\sim$25 ks for the Solid-State
Imaging Spectrometer (SIS), $\sim$28 ks for the Gas Scintillation
Imaging Spectrometer (GIS), and $\sim$2.5 ks for {\it Chandra}'s
Advanced CCD Imaging Spectrometer (ACIS). No point source was detected
at the radio position of SN1979C in a 3' diameter half power response
circle in the {\it ASCA} data. The background and galaxy subtracted SN
signal had a 3$\sigma$ upper limit to the flux of 6.3$\times 10^{-14}$
ergs s$^{-1}$ cm$^{-2}$ in the full {\it ASCA} SIS band (0.4-10.0 keV)
and a 3$\sigma$ upper limit of $<$3-4$\times$10$^{-14}$ erg s$^{-1}$
cm$^{-2}$ in the 2-10 keV band.  In the {\it Chandra} data, a source
at the position of SN1979C is marginally detected at energies below 2
keV at a flux consistent with the {\it ROSAT} HRI detection in 1995.
At energies above 2 keV, no source is detected with an upper limit of
$\sim$3$\times$10$^{-14}$ ergs s$^{-1}$ cm$^{-2}$.  These measurements
give the first ever x-ray flux limit of a Type II$_{L}$ SN above 2 keV
which is an important diagnostic of the {\it outgoing} shock wave
ploughing through the circumstellar medium.
\end{abstract}

\keywords{supernovae -- supernovae:individual (SN1979C)
  --  galaxies:individual (NGC4321)} 


\def \refhang{\noindent \hangafter=0 \hangindent=15pt }

\section{X-ray and Radio emission from Supernovae}

The observation of X-rays provides the most direct view of the region
of interaction between the ejecta from a core-collapse supernova (SN)
and the circumstellar medium of the progenitor star (see
e.g. \citealt{CF94}).  This circumstellar interaction
gives rise to copious X-ray emission within days to months of the
supernova explosion, in contrast to X-ray emission that occurs in the
later supernova remnant (SNR) during the free expansion, adiabatic, or
radiative phases.  The circumstellar medium is established by the
wind-driven mass loss from the progenitor and is heated by the
outgoing SN shock wave while the SN ejecta are heated by a reverse
shock wave.

At the same time, this interaction is expected to produce nonthermal
radio radiation; in recent years, several young SNe have been found to
be emitting in the radio bands (see \citealt{W98}). Relativistic
electrons and magnetic fields in the region, which give rise to the
radio emission, may be built up by acceleration in the shock wave and
field amplification through Rayleigh-Taylor instabilities. The
detection of strong radio emission from a SN provides good empirical
evidence of circumstellar interaction of ejecta, and such a SN is a
probable emitter at X-ray wavelengths as well. Study of the SN X-ray
and radio light curves gives an indication of how the shock-front
radius changes with time.  The latter in turn can be related to the
supernova gas radial density profile.

The observations of X-ray emission from the six supernovae discovered
prior to 1995 have been reviewed by \cite{S95}.  Currently, there are
thirteen known X-ray emitters, most of which are classified as type II
SNe on the basis of optical spectra\footnote{The 13 supernovae are:
SN1978K (II; \citealt{S99} and references therein; \citealt{R93}),
SN1979C (IIL; \citealt{Imm98a}), SN1980K (IIL; \citealt{CKF82}),
SN1986J (II; \citealt{Houck98}), SN1987A (II; \citealt{HAT96};
\citealt{Dot87}), SN1988Z (IIn; \citealt{Are99} and references
therein), SN1993J (IIb; \citealt{Zim94}), SN1994I (Ic;
\citealt{Imm98b}), SN1994W (IIn; \citealt{Sch99}), SN1995N
(\citealt{Fox00}), SN1998bw (Ic; \citealt{Pian00}), SN1999em (IIP;
\citealt{Pool2001}), and SN1999gi (IIP; \citealt{Sch01}).}.  SN1998bw (a
type Ic SN) has been associated with a gamma-ray burst source
GRB980425 and is a prototype of the collapsar model involving a
rotating massive star \citep{MW99}.  Since the predictions of the
models of GRBs regarding afterglows are different on account of
different immediate circumsource medium, it is important to detect and
monitor the X-ray emission from SNe to understand the processes
responsible for the early and long-term electromagnetic emission.  In
addition, the environments of type II and type Ib SNe are expected to
be different due to differing scenarios of mass loss from the
progenitor star.

X-ray emission from SNe that turns on shortly after the explosion and
remains active for a long time is most likely due to the thermal
emission from the hot gas produced in the interaction between the
shock wave from the explosion and the circumstellar medium.  X-ray
emission with a temperature of $\approx$ 1 keV (which falls into the
{\it ASCA} spectral band) results from a reverse shock that propagates
back into the expanding debris. The reverse shock is in turn is
developed due to the interaction of the freely expanding supernova
ejecta catching up to the decelerating (outgoing) shocked shell
\citep{CF94}.  The temperature behind the outgoing shock is typically
$>$10 keV and therefore the bulk of this radiation detected by {\it ASCA}
would appear in its higher energy bands.  In the circumstellar
interaction model, the shocked supernova ejecta is the dominant source
of X-rays because of its higher density and lower temperature.

In this paper, we report on an X-ray observation of SN1979C using {\it
ASCA} and an analysis of a {\it Chandra} observation obtained from the
data archive.  These observations provide the first constraint on the
X-ray flux in the energy band higher than 2 keV for a Type II$_{\rm
L}$ SN.

\section{Observed properties of SN1979C}

SN1979C was discovered in the optical by \citealt{J79} near maximum
optical light (m$^{max}_B \leq 12$) on April 19, 1979 in the
weakly-barred, almost face-on spiral galaxy NGC~4321 (M100).  This
galaxy is a member of the Virgo S cluster, at a distance of 17.1 Mpc
\citep{Freed94}. Three other SNe have been observed in this galaxy
this century: SN1901B, SN1914A, and SN1959E.  The J2000 position of
the SN is at $\alpha$ = $12^h 22^m 58^s.58$, $\delta$ = $+ 15^{o}
47^{'} 52^{s}.7$, offset to the southeast from the center of NGC~4321
by $\sim$100 arc seconds.

SN 1979C has been detected in the radio and X-ray bands and is
classified as a type II$_{\rm L}$ SN. The radio observation points to
a probable explosion date of April 4, 1979 \citep{Weil92}. The
supernova is thought to have originated from a red supergiant
progenitor with M$_{ZAMS} = 17-18 \pm 3 M_{\odot}$ \citep{VanDyk99}.
The SN shell expansion velocity near maximum light was $\sim 9000 \rm
\; km \; s^{-1}$ (see for example, \citealt{Pan80};
\citealt{Branch81}; \citealt{deV81}; \citealt{Frans84};
\citealt{Weil86, Weil91}; \citealt{FM93}.

SN1979C was observed by the Einstein HRI three times: in 1979 June
($\sim$4 ks), 1979 December ($\sim$13 ks), and 1980 July ($\sim$24
ks).  The upper limits for these observations, in the 0.1-4.5 keV band
and assuming a 5 keV Raymond-Smith plasma, are: June: $L_x \leq 1.8
\times 10^{40} \rm \; erg \; s^{-1}$; December: $L_x \leq 7.6 \times
10^{39} \rm \; erg \; s^{-1}$; and July: $L_x \leq 6.91 \times 10^{39}
\rm \; erg \; s^{-1}$ at a 3$\sigma$ confidence level and calculated
for a distance of 17.1 Mpc \citep{Pal81}.  An analysis of the combined
Einstein data, with a total duration $\sim$41 ks between days 64 and
454 after the explosion, gave an upper limit of 5.9 $\times 10^{39}
\rm erg \; s^{-1}$ \citep{Imm98a}. The foreground column density of
neutral hydrogen is N$_H = 2.3 \times 10^{20} \rm cm^{-2}$ (using the
dust maps of \cite{SFD98} and the E$_{\rm B-V}$-N$_{\rm H}$ relation
of \cite{PS95}).

An X-ray observation by the {\it ROSAT} HRI in 1995 June detected a
source coincident with the position of SN1979C for the first time,
more than 16 years after the explosion (source H25, \citealt{Imm98a}).
The {\it ROSAT} HRI position is $\alpha = 12^h 22^m 58^s.57$, $\delta
= + 15^{o} 47^{'} 53.5^{"}$ (J2000), within about 2.7$\arcsec$ with
the position determined by radio interferometry \citep{Pen80}.  The
measured count rate is $6.9 \times 10^{-4} \rm \; cps$. The
corresponding 0.2 - 2.4 keV flux and luminosity are $F_x = 2.9 \times
10^{-14} \rm \; erg \; cm^{-2} \; s^{-1}$ and $L_x = 1.0 \times
10^{39} \rm \; erg \; s^{-1}$ respectively, assuming a 5 keV thermal
spectrum.

\section{Observations}
\subsection{{\it ASCA}}

NGC 4321 was observed by {\it ASCA} on December 18-19, 1997.  The
Solid State Imaging Spectrometers (SIS) had useful exposure times of
24.3 ks (SIS0) and 25.2 ks (SIS1); the Gas Scintillation Imaging
Spectrometers (GIS) had exposure times of 27.5 ks (GIS2) and 28.1 ks
(GIS3).

The nominal pointing position, $\alpha$ = $12^h 23^m 58^s$, $\delta$ =
$+ 15^{o} 51^{'}$, placed the galaxy and the supernova towards the
center of the best calibrated SIS chips (chip 1 of SIS0 and chip 3 of
SIS1).

A combined SIS0+SIS1 image is shown in Figure 1.  A 3$^{\arcmin}$
diameter half power response circle is shown superposed at the
location of SN1979C. The circle approximately matches the size of the
D$_{25}$ cicle.  Emission associated with the known agglomeration of
sources in the central region of NGC~4321 is clearly seen.  On the
other hand, no point source is evident in the circle.  A point source
detection task failed to detect SN1979C in either the SIS or GIS
image.

Placing an upper limit on the {\it ASCA} band flux from SN1979C is
complicated by contamination from the integrated flux from the sources
in the nuclear region of NGC 4321.  The center of NGC 4321 is only 100
arc seconds from SN1979C, resulting in a non-negligible overlap of
point responses.  Furthermore, the emission from the galaxy is clearly
broader than that produced by a single unresolved source.  Thus
determining a count rate from SN1979C requires subtraction both of
diffuse background and the contribution of the integrated flux from
the other sources in the galaxy.

The background subtracted counts in the defined extraction circles
around the galaxy and SN are 178 $\pm 18$ counts and 82 $\pm 15$
counts respectively.  As the overlap between the two circles centered
about the galaxy and SN is estimated to be $\sim$0.4, the background-
and galaxy-subtracted SN signal is 11 $\pm 16$ counts in the full 0.4
- 10 keV band.  The implied 3$\sigma$ upper limit of the count rate is
2.0 $\times 10^{-3}$ cps. A similar analysis of SIS1 data of 25 ks
exposure, combined with the above 24 ks exposure for SIS0, gives a
combined upper limit to the count rate of 1.2 $\times 10^{-3}$
cps. The corresponding upper limit to the combined data from GIS2 and
GIS3 (28 ks each) gives 8 $\times 10^{-4}$ cps.

The net exposures, total galaxy- and background-subtracted net source
counts and their 1 $\sigma$ errors together with the 3$\sigma$ upper
limits on the SN count rates for SIS (S0 and S1) and GIS (G2 and G3)
are given in Table 1.

The above SIS combined upper limit corresponds to a flux limit of $6.3
\times 10^{-14} \; \rm erg \; cm^{-2} \; s^{-1} $ in the 0.4-10.0 keV
band for a thermal bremsstrahlung spectrum with T = 3 keV ($N_H = 2.3
\times 10^{20} \rm cm^{-2}$).  For kT = 5 keV, the flux limit is
$\sim$30\% higher, $\sim$8.2 $\times$ 10$^{39}$ erg s$^{-1}$.

In Table 1 we also list the detected x-ray source fluxes (or the upper
limits) for the previous observations obtained with {\it Einstein} and
{\it ROSAT} along with their respective band-passes and luminosities
at different epochs.

\section{{\it Chandra}}

The ACIS observation was obtained on 1999 November 6 (observation id =
400; PI=G. Garmire) as part of a survey of low-luminosity galaxy
nuclei \citep{Ho2001}.  The total exposure time was 2.5 ksec.  The
nucleus was centered on the back-illuminated (S3) chip; each chip is
8$'$ by 8$'$ in size, so the exposure, while short, potentially
captures emission from SN1979C.

Figure~\ref{Chan_fig} shows the image of the events in the 0.4-2.4 keV
band, approximating the {\it ROSAT} band.  The nucleus is clearly
visible.  Using a circle of radius 3 pixels, which encloses $\sim$92\%
of the flux, and centered on the position of SN1979C, 21$\pm$7 counts
are extracted.  The corresponding flux and luminosity, for an adopted
3 keV bremsstrahlung spectrum are 4.6$\times$10$^{-14}$ erg s$^{-1}$
cm$^{-2}$ and 1.6$\times$10$^{39}$ erg s$^{-1}$, respectively.  These
values are comparable to the values obtained by \cite{Imm98a} from
their {\it ROSAT} HRI observation.  While formally the ACIS
measurement shows an increase of a factor $\sim$2 above the {\it
ROSAT} result, that measurement requires confirmation with a longer
exposure.

Only an upper limit is possible in the 2-10 keV band; the upper limit
is $\sim$1.1$\times$10$^{39}$ erg s$^{-1}$ (3${\sigma}$).  This upper
limit is similar to the limit on the nucleus reported in
\cite{Ho2001}. A longer exposure is necessary to establish whether
SN1979C is deficient in X-rays above 2 keV.

\section{Discussion}

The {\it ROSAT} HRI flux from the measurement of \citealt{Imm98a} in
the 0.5-2.0 keV band (for the same spectrum and absorption column for
SN1979C) is $2 \times 10^{-14} \rm erg \; cm^{-2} \; s^{-1}$ which
corresponds to a luminosity of $\sim$7$\times$10$^{38}$ ergs s$^{-1}$.
The {\it ASCA} and {\it Chandra} flux limits are
$\sim$3$\times$10$^{-14}$ erg s$^{-1}$ cm$^{-2}$ or a luminosity of 1
$\times$ 10$^{39}$ erg s$^{-1}$.  For the energy bands in common
between the two satellites ($\sim$0.5-2.0 keV), {\it ROSAT} provides
the more sensitive measure; above 2 keV, a longer exposure using {\it
Chandra} will clearly reduce the upper limit.

Until now almost all SNe detected in the X-ray were measured in the
lower energy band (0.1 - 2.0 keV) of ROSAT.  Our reported measurement
of SN1979C here gives the flux limit in a harder energy band
($\gtrsim$ 2 keV) for the first time for a Type II$_{L}$ SN (the only
other X-ray detected Type II$_{L}$ is SN1980K).  The outgoing shock
wave expanding through circumstellar material creates hotter emission
compared to the reverse shock wave.  The measurement of SN X-ray
fluxes in higher energy bands constitutes an important goal.

Most extragalactic SNe detected so far in the radio are extremely
bright compared to galactic supernova remnants. For example, although
SN1970G is fainter than about half a dozen radio supernovae (some of
which like SN1988Z and SN1986J are nearly 20-30 times as bright at 6
cm), it has a peak spectral luminosity at these wavelengths that is
nearly an order of magnitude greater than the brightest galactic
supernova remnant, Cassiopeia A \citep{Weil89}.  Strong radio emission
from a supernova provides good empirical evidence of circumstellar
interaction and such a supernova is a probable emitter at X-ray
wavelengths as well.  One expects strong X-ray emission from SN1979C.

To estimate the expected X-ray fluxes from SN1979C on the basis of its
radio emission, we scaled its radio flux density (8.3 mJy) with
respect to SN1980K (another type II-L SN), as reported by
\citealt{Weil89}.  The ratio of the respective 6 cm radio flux
densities of SN1979C and SN1980K (the latter SN's radio flux is taken
from Weiler et al. as 2.6 mJy) is multiplied by the absorption-
corrected Einstein X-ray flux of 0.03 $\mu$Jy at 1 keV as reported by
\citealt{CKF82} for SN1980K, to produce an expected x-ray flux of 1.1
$\times 10^{-13} \rm erg \; cm^{-2} \; s^{-1}$ in the (0.5- 2.0) keV
band for SN1979C.

Both the {\it ROSAT} detection of the x-ray emission as well as the
{\it ASCA} and {\it Chandra} upper limits lie below this expected
value.  This lower than expected $L_x / L_R$ may mean that the X-ray
and radio emission properties of SNe can be quite variable from one
environment to another.  In other words, the emissions in the two
bands are not directly scalable in a universal way.  Or the X-ray
emission could be more strongly time variable than the radio emission.
Since the X-ray emission from a supernova is found to be primarily
from the reverse shock region, the variable outer supernova density
profile can affect the X-ray luminosity, but is unlikely to affect the
radio luminosity to the same extent.

The detection probability of a SN against a bright galactic background
can be limited by a given telescope's half power diameter, especially
when a SN fades with time. The higher spatial resolution of {\it
Chandra} improves the detection capability in this regard and is
important for their higher energy sensitivity.

\section{Acknowledgements}

The authors explicitly thank the referee for one of the speediest
reports any of us have received.  The authors are grateful to the
members of the {\it ASCA} team.  One of us (A.R.) was a Senior
Research Associate of the National Research Council at NASA/Goddard
Space Flight Center during the course of this work and acknowledges
support from the NRC and {\it ASCA} Guest Observer Program.  He also
thanks the Aspen Center for Physics for hospitality duringa portion of
this project.  The research of EMS was supported by contract number
NAS8-39073 to SAO.

\newpage

\begin{table*}[t]
\begin{center}
\caption{Observations of SN 1979C in NGC 4321}
\label{T:aray:1}
\begin{tabular}{ccccccc}
\hline
\hline
           &      & ExpT    & Band & Net source & Counts/sec  \\
 Instrument& Date &  (ksec) & (keV) & Counts & (3$\sigma$ lim) & Flux\\
\tableline
\cr
 \multicolumn{7}{c}{New Observations} \\
{\it ASCA} SIS & 1997 Dec & 24 + 25 & 0.4-10.0 & $29.6 \pm 20$ & $\leq 1.2 \times 10^{-3}$ & $\leq$6.3$\times$10$^{-14}$ \\
 (S0 + S1)  &     &  & 2.0-10.0 & $13.3 \pm 11$ & $\leq 6.7 \times 10^{-4}$ & $\leq$3.5$\times$10$^{-14}$\\

{\it ASCA} GIS & 1997 Dec & 28 + 28 & 1.0-10.0 & $-3.5 \pm 16$ & $\leq 8.4 \times 10^{-4}$ & $\leq$4.4$\times$10$^{-14}$ \\
 (G2 + G3) &        &  & 2.0-10.0 & $-12.0\pm 11$ & $\leq 6.2 \times 10^{-4}$& $\leq$3.2$\times$10$^{-14}$ \\
{\it Chandra} ACIS   & 1999 Nov  & 2.5 & 0.3-2.5 & $21{\pm}7$ & ${\sim}8{\times}10^{-3}$ & ${\sim}4.6{\times}10^{-14}$ \\
                     &           &    & 2.0-10.0 & $6.6{\pm}7.7$  & $\leq 2.89 \times 10^{-3}$ & $\leq 2.9 \times 10^{-14}$ \\ 
\cr
\multicolumn{7}{c}{Previous Observations} \\
Einstein HRI & Dec 1980 & 41 &  0.1 - 4.5 & $\cdots$ & $\cdots$ & $\leq 1.7 \times 10^{-13}$ \\

ROSAT HRI & Jun 1995 & 42.8 &  0.1 - 2.4 & $\cdots$ & $\cdots$ & $2.0 \times 10^{-14}$ \\
\tableline
\end{tabular}

Note: Flux estimated using absorbed bremsstrahlung spectrum with
kT$_{\rm Br}$ = 3 keV and column density of N$_{\rm H}$ =
3$\times$10$^{20}$ cm$^{-2}$.
\end{center}
\end{table*}

\newpage

\clearpage

\begin{figure}[t] 
\begin{center}
\scalebox{0.4}{\includegraphics{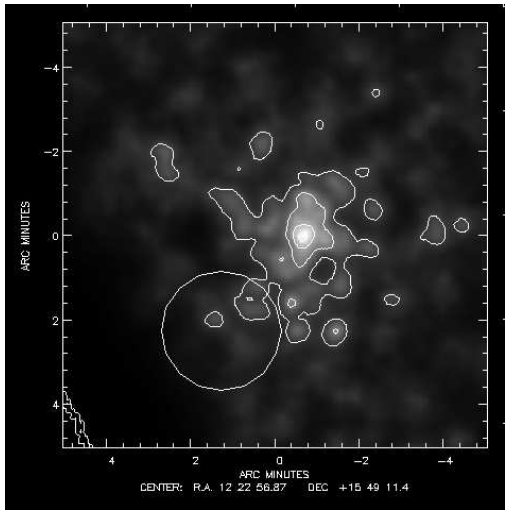}}
\vspace{5pt}
\caption{{\it ASCA} SIS image of the galaxy NGC4321 (M100) containing
the SN1979C.  The 3$'$ (diameter) extraction circle used for the
SN1979C is shown centered on its position.  SN1979C lies $\sim$100$''$
SE of the nucleus.}
\label{F:aray:1}
\end{center}
\end{figure}

\clearpage

\begin{figure}
\caption{{\it Chandra} ACIS image of NGC 4321.  The nucleus is
detected below $\sim$2 keV; the small circle indicates the location of
SN1979C.  The extracted counts at SN1979C's location are consistent
with the expected flux from the {\it ROSAT} HRI measurement.}
\label{Chan_fig}
\scalebox{0.4}{\rotatebox{-90}{\includegraphics{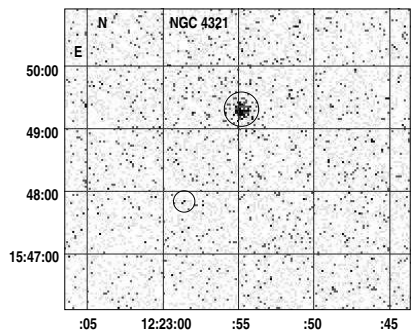}}}
\end{figure}

\end{document}